\documentclass[a4paper,11pt]{article}
\usepackage{amsfonts,amssymb}
\usepackage{theorem}
\usepackage{epsf}
\def\G{\Gamma}

\begin{document}
\rightline{IFUM-803-FT and MPI-PhT/2004-91}
\vskip 2.7 truecm
\Large
\bf
\centerline{Some Conjectures on the Limit }
\par
\centerline{ of Infinite Higgs Mass}
\normalsize
\rm
\vskip 1.2 truecm
\large
\centerline{Ruggero Ferrari $^a$
\footnote{E-mail address: {\tt ruggero.ferrari@mi.infn.it}},
Marco Picariello $^a$
\footnote{E-mail address: {\tt marco.picariello@mi.infn.it}}
and Andrea Quadri $^b$ \footnote{E-mail address: {\tt quadri@mppmu.mpg.de}}}
\vskip 0.4 truecm
\normalsize
\centerline{$^a$Phys. Dept. University of Milan, 
via Celoria 16, 20133 Milan, Italy } 
\centerline{I.N.F.N., sezione di Milano} 
\centerline{$^b$Max-Planck-Institut f\"ur Physik (Werner-Heisenberg-Institut)} 
\centerline{F\"ohringer Ring, 6 - D80805 M\"unchen, Germany}
\vskip 2.0  truecm
\normalsize
{\bf
\centerline{Abstract}}
\vskip 0.4 truecm
We consider a possible field theory candidate for the electroweak
$SU(2)\otimes U(1)$
model where the limit of infinitely sharp  Higgs potential
is performed. We show that it is possible
to formulate such a limit as a St\"uckelberg massive
non Abelian gauge theory.
\rm
\begin{quotation}

\end{quotation}

\newpage

\section{Introduction}
\label{sec:intr}
The electroweak model \cite{ew} as a part of the Standard Model has had
impressive confirmations \cite{SM} and great is the
expectation for the experimental evidence of the last important part:
the Higgs sector. The confirmation  of the
spontaneous breaking of the $SU(2)\otimes U(1)$ via Higgs
mechanism would be an important step in the development
of quantum field theory. Gauge theories have proved to be the
correct framework for the description of the elementary particle
world. Their implementation by means of the spontaneously broken symmetry 
mechanism is
then expected as a further development.
\par
This satisfactory situation is not a reason to stop searching
for new implementations of gauge theories. In particular
we are interested in the formulation of massive gauge theories
\`a la St\"uckelberg, where the mass for the gauge field is
introduced without breaking the local gauge
invariance (BRST invariance in the quantum version of the theory).
Of course we face a presently insurmountable objection on the 
non-renormalizability of the perturbation theory. However many options
could in principle be at disposal as a way out to this objection,
as lattice calculation of non perturbative effects.
\par
In the present paper we look for a formulation of the electroweak
$SU(2)\otimes U(1)$
model where the masses of the vector mesons are introduced via the
St\"uckelberg method. The existence of such a formulation
is important for two reasons. First it opens the way to a possible
use of the equivalence theorem in order  to set a bridge with
the unitary gauge of massive gauge theory. Second it would allow
topological arguments on the ew model based on the fact that
the St\"uckelberg mass makes use of the flat connection built on
a non linear sigma model.
\par
We imagine our final model as a limit for infinitely sharp Higgs
potential (in standard notations $\lambda \to \infty$) without,
however, committing ourselves with a limit on the mass of
the Higgs boson. The two problems seem to us very distinct and
only a reasonable theory of the non-linear sigma model would
be able to shed some light on their inter correlations.
\par
The formal limit of $\lambda \to \infty$ has been considered by
some authors \cite{limit} and there is unanimous consensus that the
boson sector is described by a non-linear sigma model. In the
present paper we show that the scalar field can be accommodated
so that it appears only via a flat connection, i.e. a pure-gauge
field.
\par
The plan of the paper is somehow reversed. We prefer to construct
a model which has no physical interpretation, but serves to us
to put down the rational for the construction of the model we are really
interested in. Our guide in the construction will simply be
the compatibility of the main features of the present phenomenology with the tree 
approximation of the  theory.
\section{Preliminary considerations}
\label{sec:prel}
Let us consider a fully globally symmetric $SU(2)$ with
a mass term in the Proca gauge
\begin{eqnarray} 
S= \int d^4 x \left(
-\frac{1}{4}G_{a\mu\nu}G_a^{\mu\nu} + m^2 ~Tr~\left[
 A_\mu A^\mu  \right]
\right),
\label{prel.1}
\end{eqnarray}
where
\begin{eqnarray}
 A_\mu = \frac{1}{2}\tau_aA_{a\mu }.
\label{prel.2}
\end{eqnarray}
$\tau_a$ are the Pauli matrices.
Let us now perform a formal operator valued local $SU(2)$ 
transformation in order
to introduce the St\"uckelberg field
\begin{eqnarray}
A'_\mu = \Omega^\dagger A_\mu \Omega 
-\frac{ i}{g} \Omega^\dagger\partial_\mu \Omega,
\label{prel.3}
\end{eqnarray}
with the constraint
\begin{eqnarray}
\Omega \in SU(2) \quad \Longrightarrow \quad
\Omega^\dagger = \Omega^{-1},\quad \det\Omega=1.
\label{prel.3.1}
\end{eqnarray}

One gets
\begin{eqnarray}&& 
S= \int d^4 x \Big\{
-\frac{1}{4}G_{a\mu\nu}G_a^{\mu\nu} 
\nonumber \\&&
+\frac{m^2}{g^2}  ~Tr~\left[
\left(g\Omega^\dagger A_\mu \Omega  - i \Omega^\dagger\partial_\mu\Omega
\right)\left(g
\Omega^\dagger A^\mu  \Omega - i \Omega^\dagger\partial^\mu\Omega
\right)\right] 
\Big\}.
\nonumber \\
\label{prel.4}
\end{eqnarray}  
{\bf Proposition 1:} {\sl
Each element of the matrix
(each term in round brackets in eq. (\ref{prel.4}))
\begin{eqnarray}
\left(g\Omega^\dagger A_\mu \Omega  - i \Omega^\dagger\partial_\mu\Omega
\right)_{ab}
\label{prel.4.1}
\end{eqnarray}  
 is invariant under local $SU(2)$ 
left transformations
\begin{eqnarray}&&
A'_\mu = U_L A_\mu U_L^\dagger -\frac{ i}{g} U_L\partial_\mu U_L^\dagger
\nonumber \\&&
\Omega' = U_L\Omega.
\label{prel.5}
\end{eqnarray}
}
In fact
\begin{eqnarray}&&
g\Omega^{'\dagger} A'_\mu \Omega'  - i \Omega^{'\dagger}\partial_\mu\Omega'
\nonumber \\&&
= g\Omega^\dagger A_\mu \Omega 
-i\Omega^\dagger U_L^\dagger U_L\partial_\mu U_L^\dagger U_L\Omega
- i \Omega^\dagger U_L^\dagger\partial^\mu(U_L\Omega)
\nonumber \\&&
= g\Omega^\dagger A_\mu \Omega - i \Omega^\dagger\partial^\mu\Omega.
\label{prel.6}
\end{eqnarray}
\par\noindent As a consequence of the above result,
we can construct an arbitrary number of  invariants since there are no
constraints on the dimensionality of the coupling constants. Among them
some are bilinear in the gauge field. For
instance the following two terms
\begin{eqnarray}&&\int d^4 x 
\frac{m'^2}{g^2} v^\dagger \Big[ 
\left(g\Omega^\dagger A_\mu \Omega  - i \Omega^\dagger\partial_\mu\Omega
\right)^2
\Big]v,
\label{prel.7a}
\\&&
\int d^4 x 
\frac{m'^2}{g^2}\Big(  v^\dagger 
\left(g\Omega^\dagger A_\mu \Omega  - i \Omega^\dagger\partial_\mu\Omega
\right)v
\Big)^2,
\label{prel.7b}
\end{eqnarray}
where $v$ is any constant spinor which we choose to normalize by
\begin{eqnarray} 
v^\dagger v = 1.
\label{prel.8}
\end{eqnarray}
\par The proposition 1. can be extended to $SU(2)\otimes U(1)$.
One enlarges the  local gauge transformations in eq. (\ref{prel.5}) 
by introducing a further abelian gauge field \cite{Sonoda:1983kh}
\begin{eqnarray}&&
B'_\mu = B_\mu - \frac{ 1}{g'} \partial_\mu \lambda
\nonumber \\&&
A'_\mu = U_L A_\mu U_L^\dagger -\frac{ i}{g} U_L\partial_\mu U_L^\dagger
\nonumber \\&&
\widehat\Omega' = e^{i\lambda} U_L\widehat\Omega
\nonumber \\&&
 U_L\in SU(2) .
\label{prel.8.2}
\end{eqnarray}
The transformation properties of $\widehat\Omega$ implies that
four real fields are necessary in order to describe the degree
of freedom
\begin{eqnarray}
\widehat\Omega = \exp(i\phi_B)(\phi_0 +i \tau_a\phi_a), \qquad
\phi_0^2 + \vec\phi ^2 =1.
\label{prel.8.3}
\end{eqnarray}
\noindent {\bf Proposition 2:} {\sl
 Each element of the following
matrix 
\begin{eqnarray}
\left(g'B_\mu+g\widehat\Omega^\dagger A_\mu \widehat\Omega  
- i \widehat\Omega^\dagger\partial_\mu\widehat\Omega
\right)_{ab}
\label{prel.8.4}
\end{eqnarray}
is invariant under the transformations (\ref{prel.8.2}).
}
Thus, again, any Lorentz invariant function of (\ref{prel.8.4}) is a possible
term of the action. In particular the action for eq. (\ref{prel.7a})
becomes
\begin{eqnarray}&& 
S= \int d^4 x \Big(
-\frac{1}{4}G_{a\mu\nu}G_a^{\mu\nu}  -\frac{1}{4}F_{\mu\nu}F^{\mu\nu}
\nonumber \\&&
+\frac{m^2}{g^2}  ~Tr~\left\{
\left(g'B_\mu+g\widehat\Omega^\dagger A_\mu \widehat\Omega  
- i \widehat\Omega^\dagger\partial_\mu\widehat\Omega
\right)\left(g'B^\mu+g
\widehat\Omega^\dagger A^\mu  \widehat\Omega 
- i \widehat\Omega^\dagger\partial^\mu\widehat\Omega
\right)\right\} 
\Big).
\nonumber \\
\label{prel.8.5}
\end{eqnarray}
But other terms are possible \cite{Sonoda:1983kh}, as, for instance, 
from eq. (\ref{prel.7b})
\begin{eqnarray}&& 
S= \int d^4 x \Big(
-\frac{1}{4}G_{a\mu\nu}G_a^{\mu\nu}  -\frac{1}{4}F_{\mu\nu}F^{\mu\nu}
\nonumber \\&&
+\frac{m^2}{g^2}  \left\{v^\dagger
\left(g'B_\mu+g\widehat\Omega^\dagger A_\mu \widehat\Omega  
- i \widehat\Omega^\dagger\partial_\mu\widehat\Omega
\right)v \right\} ^2
\Big).
\nonumber \\
\label{prel.8.6}
\end{eqnarray}
Clearly the St\"uckelberg theory based on the gauge transformations
in eq. (\ref{prel.8.2}) is not a good candidate as a limit of the
ew model. First $\widehat\Omega$ contains a $U(1)$ factor which
is really not required. Second, due to the freedom of introducing
more terms bilinear in the vector fields, the  St\"uckelberg mass
formulation cannot reproduce, even at the tree level, the
phenomenology electroweak interactions (as the neutral currents
with the correct couplings or the $\rho$ parameter).

\section{Extension to $SU(2)\otimes SU(2)$}
\label{sec:full}

Let us consider again  the action (\ref{prel.4}). The form 
of the mass term suggests that one can enlarge the symmetry
of the action by  considering
the right $SU(2)$ global transformations. This model has been studied at
length by Bardeen and Shizuya in Ref. \cite{limit}.
One can extend the symmetry to the right $SU(2)$ local transformations
by introducing  a new  field such that
\begin{eqnarray}&&
B'_\mu=  U_R B_\mu  U_R^\dagger 
-\frac{ i}{g'}\partial_\mu  U_R  U_R^\dagger
\nonumber \\&&
A'_\mu = U_L A_\mu U_L^\dagger -\frac{ i}{g} U_L\partial_\mu U_L^\dagger
\nonumber \\&&
\Omega' =  U_L\Omega{ U_R^\dagger}.
\label{full.9}
\end{eqnarray}
Then under local $SU_L(2)\otimes SU_R(2)$ transformations
we have
\begin{eqnarray}
\left(g'B_\mu + g
\Omega^\dagger A^\mu  \Omega - i \Omega^\dagger\partial^\mu\Omega
\right) \to  U_R \left(g'B_\mu + g
\Omega^\dagger A^\mu  \Omega - i \Omega^\dagger\partial^\mu\Omega
\right) U_R^\dagger.
\label{full.10}
\end{eqnarray}
Consequently the only term invariant under $SU_L(2)\otimes SU_R(2)$
local transformations, that is bilinear in the gauge fields, is
\begin{eqnarray}&&
\int d^4 x 
\frac{m^2}{g^2}
\nonumber \\&&  
Tr~\left\{
\left(g' B_\mu + 
g\Omega^\dagger A_\mu \Omega  - i \Omega^\dagger\partial_\mu\Omega
\right)\left(g' B_\mu +g
\Omega^\dagger A^\mu  \Omega - i \Omega^\dagger\partial^\mu\Omega
\right)\right\} .
\nonumber \\
\label{full.11}
\end{eqnarray}
We can add chiral fermions to this theory. The necessity to introduce
a Yukawa coupling 
\begin{eqnarray}
\bar\psi_L \Omega\psi_R + \bar\psi_R \Omega^\dagger\psi_L
\label{full.12}
\end{eqnarray}
determines the transformation properties of the fermion fields
and therefore the kinetic part of the action
\begin{eqnarray}
\int d^4 x \Big [\bar\psi_L \gamma^\mu \left(i\partial_\mu + g A_\mu
\right)\psi_L
+\bar\psi_R \gamma^\mu \left(i\partial_\mu + g' B_\mu
\right)\psi_R
\Big].
\label{full.13}
\end{eqnarray}
This model has three massive  and three massless vector mesons
at the tree level of the perturbation theory. 
In order to get a
reasonable model, one has to devise a mechanism that removes two
of the massless vector mesons. 
Moreover the correct quantum
numbers of quark and leptons can be obtained by introducing
a novel $U(1)_{L-B}$ symmetry in analogy to celebrated left-right
symmetric models \cite{l-r}.

In this line of thought 
a Higgs field in the adjoint representation of $SU_R(2)$ can be introduced 
in order to generate 
a mass term for two of the three vector mesons $B_\mu^a$, keeping 
the third one massless. The massive $SU_R(2)$-gauge bosons as well as the adjoint
Higgs field  are to be made sufficiently heavy in order to reproduce the 
low-energy Standard Model phenomenology.

\medskip
In a somehow different fashion an extended symmetry content  
can be obtained 
by gauging the hidden symmetry of non-linear $\sigma$ model~\cite{bess}.
This results in an extra local $SU(2)$ invariance, which is 
implemented by means of an additional set of heavy gauge bosons.
Along these lines  
a phenomenologically viable extension of the SM
has been derived in Refs.~\cite{bess}.

\section{$SU(2)\otimes U(1)$ Symmetry }
\label{sec:broken}
The toy model of the Section \ref{sec:full} suggests a way
to build a St\"uckelberg theory of the ew model which is
not in contrast with the gross features of phenomenology.
Let us consider the reduction of the full symmetry
$SU_L(2)\otimes SU_R(2)$ to  $SU_L(2)\otimes U(1)$. 
When we consider this reduction on the Fermion sector
one cannot forget that left-right symmetric models necessitate
of an extra $U(1)$ invariance in order to distinguish leptons from
quarks, as mentioned at the end of Section \ref{sec:full}. Then
the process is not simply the reduction of $SU_R(2)\to U_R(1)$.
However
when we consider the gauge sector by itself, then this reduction
is achieved simply by $SU_R(2)\to U_R(1)$ i.e. by imposing
invariance under the transformations (see eq. (\ref{full.9}))
\begin{eqnarray}&&
\tau_3 B'_\mu=  \tau_3 B_\mu
-\frac{ i}{g'} \partial_\mu U_R  U_R^\dagger
\nonumber \\&&
A'_\mu = U_L A_\mu U_L^\dagger -\frac{ i}{g} U_L\partial_\mu U_L^\dagger
\nonumber \\&&
\Omega' =  U_L\Omega{ U_R^\dagger},
\label{broken.1}
\end{eqnarray}
with 
\begin{eqnarray}&&
U_L\in SU_L(2)
\nonumber \\&&
U_R = \exp(i\lambda \frac{\tau_3}{2})\in U(1).
\label{broken.2}
\end{eqnarray}
A reduction of the symmetry of the action allows to have more invariant
terms. We use a fixed vector, eigenvector of $\tau_3$ 
\begin{equation}
v_+ =
\left(
\begin{array}{l}
1 \\
0
\end{array}
\right)
\label{broken.3}
\end{equation}
then we have an invariant mass term described by the expression
\begin{eqnarray}&&
v_+^\dagger~
\left(g' B_\mu \frac{\tau_3}{2} + 
g\Omega^\dagger A_\mu \Omega  - i \Omega^\dagger\partial_\mu\Omega
\right)\left(g' B^\mu\frac{\tau_3}{2} +g
\Omega^\dagger A^\mu  \Omega - i \Omega^\dagger\partial^\mu\Omega
\right)v_+
\nonumber \\&&
=\Phi^\dagger \left\{\left(
\frac{g'}{2} B_\mu + g A_\mu  + i \stackrel{\leftarrow}{\partial}_\mu
\right)\left(
\frac{g'}{2} B^\mu + g A^\mu - i \stackrel{\rightarrow}{\partial^\mu}
\right)\right\}\Phi
\label{broken.4}
\end{eqnarray}
where
\begin{eqnarray}
\Phi = \Omega v_+, \qquad\Phi^\dagger \Phi =1,
\label{broken.5}
\end{eqnarray}
i.e. the formal limit of $\lambda \to \infty$ in the electroweak
model.
\par
In terms of scalar fields
\begin{equation}
\Phi =
\left(
\begin{array}{l}
\phi_0 +i\phi_3 \\
i\phi_1 -\phi_2
\end{array}
\right)
\label{broken.6}
\end{equation}
An other invariant is built with 
\begin{equation}
v_- =
\left(
\begin{array}{l}
0 \\
1
\end{array}
\right)
\label{broken.7}
\end{equation}
and one gets
\begin{eqnarray}&&
v_-^\dagger~
\left(g' B_\mu \frac{\tau_3}{2} + 
g\Omega^\dagger A_\mu \Omega  - i \Omega^\dagger\partial_\mu\Omega
\right)\left(g' B_\mu\frac{\tau_3}{2} +g
\Omega^\dagger A^\mu  \Omega - i \Omega^\dagger\partial^\mu\Omega
\right)v_-
\nonumber \\&&
=\tilde\Phi^\dagger \left\{\left(-
\frac{g'}{2} B_\mu + g A_\mu  + i \stackrel{\leftarrow}{\partial}_\mu
\right)\left(-
\frac{g'}{2} B^\mu + g A^\mu - i \stackrel{\rightarrow}{\partial^\mu}
\right)\right\}\tilde\Phi
\label{broken.8}
\end{eqnarray}
where
\begin{equation}
\tilde\Phi \equiv \Omega v_- =
\left(
\begin{array}{l}
 i\phi_1 +\phi_2\\
\phi_0 -i\phi_3
\end{array}
\right).
\label{broken.9}
\end{equation}
Since 
\begin{eqnarray}
\tilde{\Phi} = \epsilon \Phi^* 
\label{broken.10}
\end{eqnarray}
where
\begin{equation}
\epsilon =
\left(
\begin{array}{rr}
0 & -1\\
1 & 0
\end{array}
\right),
\label{broken.11}
\end{equation}
it is easy to prove that the two invariants in eqs. (\ref{broken.4}) and
(\ref{broken.8}) are the same. Thus we can write eq.  (\ref{broken.4})
also in the form
\begin{eqnarray}
\frac{1}{2}~Tr ~
\left(g' B_\mu \frac{\tau_3}{2} + 
g\Omega^\dagger A_\mu \Omega  - i \Omega^\dagger\partial_\mu\Omega
\right)\left(g' B_\mu\frac{\tau_3}{2} +g
\Omega^\dagger A^\mu  \Omega - i \Omega^\dagger\partial^\mu\Omega
\right).
\label{broken.12}
\end{eqnarray}
This form  will be useful  to discuss custodial symmetry in
section \ref{sec:cs}.

\subsection{Mass terms }

Now we consider the Yukawa couplings for the Fermi-Dirac fields.
Since we have reduced the right symmetry $SU_R(2) \to U(1)$ we can
build more invariants as in eq. (\ref{full.12}).
Then the Yukawa sector can be enlarged to a two-parameter space by 
\begin{eqnarray}
f \bar u_R \Phi^\dagger\psi_L + \tilde f \bar d_R {\tilde\Phi}^\dagger\psi_L
+ h.c.
\label{mass.5}
\end{eqnarray}
which can be extended to the Cabibbo-Kobayashi-Maskawa \cite{CKM} 
theory by considering
the adequate number of Fermion families. 
The requirement that the Yukawa term in eq. (\ref{mass.5}) is invariant
under $SU_L(2)\otimes U(1)$ leaves enough freedom to reproduce the
standard internal quantum numbers of leptons and quarks. The only
constraint on the $U(1)$ charge of the Fermions comes from the balance implied
by the equation (\ref{mass.5}).
\section{Custodial symmetry and Slavnov-Taylor identities}
\label{sec:cs}
The fundamental problem of the existence of a theory
which is the limit for infinite Higgs-self-coupling cannot
avoid to exploit some general properties of the conventional
ew model. Two items are exceedingly important for somewhat
different reasons. We only sketch  them here.

\subsection{Custodial symmetry}
The Standard model has a custodial symmetry \cite{custodial} which
prevents large corrections to the $\rho$ parameter \cite{vanderBij:1983bw}. The
ew model becomes invariant under a {\sl global} $SU_R(2)$ symmetry
when the $U(1)$ coupling constant is switched off and the Yukawa
couplings are put equal within a single flavor family. The St\"uckelberg
formulation we are presenting justs hands on this symmetry. 
\par
This feature can be seen from eq. (\ref{broken.12}) in Section
\ref{sec:broken}, where for $g'=0$ and equal Yukawa couplings
this symmetry is explicitly displayed. The consequences of
this symmetry are similar to those of the ew model.
\par
By changing basis of the global  $SU_L(2)\otimes SU_R(2)$
transformations we introduce $SU_V(2)$ and $SU_A(2)$ transformations.
In particular on the St\"uckelberg field we get
\begin{eqnarray}&&
\delta_V \phi_0 = 0
\nonumber\\&&
\delta_V \phi_a =  
\frac{\delta\omega_{Vc}}{2}
\epsilon_{abc}\phi_b
\label{cs.1}
\end{eqnarray}
and
\begin{eqnarray}&&
\delta_A \phi_0 = - \frac{\delta\omega_{Aa}}{2}\phi_a
\nonumber\\&&
\delta_A \phi_a =  \frac{\delta\omega_{Aa}}{2}\phi_0.
\label{cs.2}
\end{eqnarray}
Eq.(\ref{cs.1}) corresponds to the choice $U_L=U_R$ in eq.(\ref{full.9}), 
eq.(\ref{cs.2}) to $U_L=U_R^\dagger$.
Setting $\Omega_{\rm Higgs} = \phi_0 ~ 1 + i \phi^a \tau^a$ after spontaneous 
symmetry breaking one has
$<\Omega_{\rm Higgs}>= v ~ 1$. $<\Omega_{\rm Higgs}>$ is left invariant under 
the transformation in eq.(\ref{full.9})
provided that $U_L=U_R$. This choice gives rise to the custodial symmetry 
in eq.(\ref{cs.1}), 
leaving  the field $\phi_0$ is invariant. The spontaneous
breakdown does not affect the generators of the $SU_V(2)$
group of transformations.

\subsection{Slavnov-Taylor identities}
The BRST \cite{BRST} invariance properties of the St\"uckelberg formulation 
of the electroweak model are identical to those of the standard model. 
The fact that
$\phi_0$ is a composite field does not change its transformation properties,
since $\Omega^\dagger \Omega=1$ is a gauge- (and thus BRST-)invariant 
constraint.
Thus we have for the $SU(2)\otimes U(1)$ sector
\begin{equation}
\begin{array}{ll}
s_1\, A_{a\mu} = \partial_\mu c_a -  A_{c\mu}\epsilon_{abc}c_b
&s_0\, A_{a\mu} = 0 \\
s_1\, B_\mu =0   &   s_0\, B_\mu = \partial_\mu c_0 \\
s_1\, c_a = -\frac{1}{2} \epsilon_{abc}c_b c_c & s_0\,c_a=0\\
s_1\, c_0 = 0 & s_0\,c_0 = 0 \\
s_1\, \bar c_a = b_a &  s_0\,\bar c_a =0\\
s_1\, b_a = 0 & s_0\,  b_a = 0\\
s_1\, \bar c_0 = 0 &  s_0\,\bar c_0 =b_0\\
s_1\, b_0 = 0 & s_0\,  b_0 = 0\\
s_1\, \phi_a = -\epsilon_{abc}c_b  \phi_c + c _a\phi_0 &  s_0\,\phi_1=c_0 \phi_2\\
& s_0\,\phi_2=-c_0 \phi_1\\
& s_0\,\phi_3= -c_0 \phi_0 \\
s_1\, \phi_0 = -  c _a\phi_a & s_0\,\phi_0= c_0 \phi_3 \\
.....{\rm matter~ part}&
\end{array}
\label{cs.3}
\end{equation}
where the fields $b_a,~ b_0$ are the Lagrange multipliers used to impose the
Landau gauge.
Clearly these BRST transformations lead to the same Slavnov Taylor
identities~\cite{st} valid for the Standard Model in the Electroweak sector.

The composite nature of the field $\phi_0(x)$ is reflected in
the formulation of the relevant Slavnov Taylor identities ~\cite{st}.
In order to define the correlation functions of $\phi_0(x)$
the latter has to be coupled in the tree-level
approximation  of the vertex functional $\G^{(0)}$ to the external 
source $\beta(x)$.
The external source $\beta^*(x)$
coupled to the BRST variation $s \phi_0(x)$ has also to be included.
The $\beta,\beta^*$-dependence of $\G^{(0)}$ is then
\begin{eqnarray}
\int d^4x \, \Big ( \beta(x) \phi_0(x) + \beta^*(x) s \phi_0(x) \Big ) \, .
\label{newcons.1}
\end{eqnarray}

The St\"uckelberg model is not power-counting renormalizable.
The cohomology of the relevant BRST differential in eq.(\ref{cs.3})
has been computed in \cite{Henneaux:1998hq}. It has been shown there that the most general deformation
of the action (in the space of local formal power series) is given, up to trivial invariants, by a strictly gauge invariant term plus winding number terms. The latter are irrelevant in perturbation theory.
Moreover, there is no perturbative anomaly.
As a consequence, the St\"uckelberg model turns out to be renormalizable in the modern sense
of \cite{Gomis:1995jp}. 

In order to discuss Physical Unitarity the construction of a nilpotent BRST charge $Q$ 
\cite{Curci:1976yb,Kugo:1977zq} is required.
The conditions on the operator $Q$ needed to establish Physical Unitarity have been given
under fairly general assumptions (not restricted to the perturbative
framework) in \cite{gen_unit}. They provide constraints on the quantization procedures 
aiming at a non-perturbative definition of the St\"uckelberg model.
Under the assumption that the subtraction scheme fulfills the ST identities, 
the cancellation mechanisms implied by perturbative Physical Unitarity 
have been analyzed in \cite{Ferrari:2004pd}.

\section{Other invariants}

The St\"uckelberg mass term in eq. (\ref{broken.12}) has been
constructed as an effective field theory 
in the spirit of mantaining most of the properties
of the Standard model as it has been shown in Section \ref{sec:cs}.
This requirement is important if one hopes to establish some
relation between the Higgs formulation and the  St\"uckelberg's one.

\par
There is a further invariant under the {\sl local } $SU(2)\otimes U(1)$
which is also a bilinear form in the vector fields \cite{Peccei:1989kr}.
This form can be constructed by noticing that
\begin{eqnarray}
v_+^\dagger~
\left(g' B_\mu \frac{\tau_3}{2} + 
g\Omega^\dagger A_\mu \Omega  - i \Omega^\dagger\partial_\mu\Omega
\right)
v_-
\label{other.1}
\end{eqnarray}
under the transformations in eq. (\ref{broken.1}) becomes
\begin{eqnarray}
\exp(i\lambda)~
v_+^\dagger~
\left(g' B_\mu \frac{\tau_3}{2} + 
g\Omega^\dagger A_\mu \Omega  - i \Omega^\dagger\partial_\mu\Omega
\right)
v_-
\label{other.1.1}
\end{eqnarray}
Then the following bilinear form in the gauge field is invariant
\begin{eqnarray}&&
v_+^\dagger~
\left(g' B_\mu \frac{\tau_3}{2} + 
g\Omega^\dagger A_\mu \Omega  - i \Omega^\dagger\partial_\mu\Omega
\right)
v_-
\nonumber \\&& v_-^\dagger
\left(g' B_\mu\frac{\tau_3}{2} +g
\Omega^\dagger A^\mu  \Omega - i \Omega^\dagger\partial^\mu\Omega
\right)v_+
\nonumber \\&&
=\left\{\Phi^\dagger
 \left(g A_\mu  + i \stackrel{\leftarrow}{\partial}_\mu
\right)\tilde\Phi\right\}~
\left\{\tilde\Phi^\dagger
\left(g A^\mu - i \stackrel{\rightarrow}{\partial^\mu}
\right)\Phi\right\}.
\label{other.2}
\end{eqnarray}
This term can be dismissed only on the basis of the requirement
that a custodial symmetry is present as discussed in Section \ref{sec:cs}.
In the standard model it is not present since it contains couplings
that cause non renormalizability of the theory.

\section{Conclusions}
In the present paper we have shown that the formal limit of infinite
Higgs potential ($\lambda\to\infty$) can be casted in a theory with
a St\"uckelberg mass. Moreover, always at the formal level, the
proposed limit
enjoys the same custodial symmetry and the BRST invariance properties
as the Standard Electroweak model. We make clear that our work
does not prove the existence of such a limit and, if such a limit exists,
that no Higgs boson is present. In fact a physical boson particle
might show up in many ways, e.g. as a  non perturbative effects. 
Nevertheless it is rather surprising that such
a powerful tool as the Slavnov-Taylor identities can be traced 
also in the limit
of infinite Higgs potential.
\section*{Acknowledgments}

One of us (RF) is honored to  gratefully
acknowledge the warm hospitality and the partial financial support 
of the Laboratoire de Physique Th\'eorique d'Orsay (Universit\'e Paris Sud)
where part of this work has been accomplished.


\bibliography{reference}

\end{document}